\documentclass[
reprint,
superscriptaddress,
amsmath,amssymb,
aps,
prl,
longbibliography,
lengthcheck
]{revtex4-1}
\usepackage[pdftex]{graphicx,color}
\usepackage{epstopdf}
\usepackage[hidelinks]{hyperref}
\usepackage{bm}

\begin{document}
	
	\title{Berry phase of phonons and thermal Hall effect in nonmagnetic insulators}
	
   \author{Takuma Saito}
   \affiliation{
   Department of Applied Physics, The University of Tokyo, Bunkyo-ku, Tokyo, 113-8656, JAPAN 
   }

	\author{Kou Misaki}
	\affiliation{
		Department of Applied Physics, The University of Tokyo, Bunkyo-ku, Tokyo, 113-8656, JAPAN 
	}

   \author{Hiroaki Ishizuka}
   \affiliation{
   Department of Applied Physics, The University of Tokyo, Bunkyo-ku, Tokyo, 113-8656, JAPAN 
   }

   \author{Naoto Nagaosa}
   \affiliation{
   Department of Applied Physics, The University of Tokyo, Bunkyo-ku, Tokyo, 113-8656, JAPAN 
   }
   \affiliation{
   RIKEN Center for Emergent Matter Science (CEMS), Wako, Saitama, 351-0198, JAPAN
   }
	
	\date{\today}
	
\begin{abstract}
A mechanism for phonon Hall effect (PHE) 
in non-magnetic insulators under an external magnetic field is theoretically studied. 
PHE is known in (para)magnetic compounds, 
where the magnetic moments and spin-orbit interaction play an essential role. 
In sharp contrast, we here show that a non-zero Berry curvature of acoustic 
phonons is induced by an external magnetic field due to the 
correction to the adiabatic Born-Oppenheimer approximation. 
This results in the finite thermal Hall conductivity $\kappa_H$
in nonmagnetic band insulators. 
Our estimate of $\kappa_H$ for a simple model gives 
$\kappa_H \sim 1.0\times 10^{-5} $[W/Km] at $ B=10 $[T] and $ T=150 $[K].
\end{abstract}
	
\maketitle
	
\textit{Introduction} --- Hall effects are one of the most important subjects in condensed matter
physics. It provides the information on the sign and density of the carriers
in semiconductors, and the shape of the Fermi surface in metals. 
Since the discovery of quantum Hall effect~\cite{cage2012quantum}, the close connection of Hall 
effects to topological nature of electronic states in solids has become a
keen issue. In addition to the quantum Hall effect, the anomalous Hall effect
in metallic magnets~\cite{nagaosa2010anomalous}, 
and spin Hall effect in semiconductors~\cite{murakami20111}
are interpreted 
as the consequence of the geometric phase of Bloch wavefunctions, i.e.,
Berry phase in solids~\cite{xiao2010berry}. Berry phase can be nonzero even for the neutral 
particles such as photons~\cite{onoda2004hall} and magnons~\cite{katsura2010theory,matsumoto2011theoretical,matsumoto2011rotational}
, and the Hall effects of these
particles are observed experimentally in Ref.~\onlinecite{hosten2008observation}  and Ref.~\onlinecite{onose2010observation}, respectively.

Phonon is another neutral particle in solids, and the thermal Hall effect of 
phonons, phonon Hall effect (PHE), 
has been studied experimentally~\cite{strohm2005phenomenological, inyushkin2007phonon}
and theoretically~\cite{sheng2006theory,kagan2008anomalous,wang2009phonon,zhang2010topological,agarwalla2011phonon,kronig1939r,qin2012berry}. 
In most of the theoretical works, the Raman-type interaction is assumed
whose Hamiltonian reads
\begin{align}
	H_{\textrm{Raman}} = \lambda \bm{M} \cdot (\bm{u} \times \bm{P}).
\label{Raman}
	\end{align}
where $\bm{M}$ is the electronic magnetization, 
$\bm{u}$ the displacement of nucleus, and 
$\bm{P}$ the momentum of nucleus. This coupling $\lambda$
is supposed to originate from the spin-lattice interaction, but the
microscopic theory for $\lambda$ is missing in most of the cases.  

The charge of the phonon, however, is a subtle issue
because the atomic nuclei are positively charged, which is compensated
by the electrons. Then the screening effect of electrons should be 
treated properly to ensure the neutral nature of phonons.
The conventional formalism to study the electron-phonon coupled system 
is the Born-Oppenheimer (BO) approximation~\cite{mead1992geometric}, which uses the fact 
that the electron mass $m$ is much lighter than that of atoms $M$. 
Writing the wave function as the product of electronic and 
nuclear part, i.e., 
 $\Psi(r, R)= \psi_{\textrm{el}} (r, R) \phi_{\textrm{nucl}}(R)$
with $r$ and $R$ being the
position of electrons and nuclei, respectively,
the ratio of the length scales $\ell_{\textrm{el}}$ and $\ell_{\textrm{nucl}}$ for 
$\psi_{\textrm{el}} (r, R)$ and $\phi_{\textrm{nucl}}(R)$ is estimated as
\begin{align}
	\ell_{\textrm{el}}/\ell_{\textrm{nucl}} \sim (m_{\textrm{nucl}}/m_{\textrm{el}})^{1/4}.
	\label{BO}
\end{align}
Therefore, the derivative $\nabla_R$ on 
$\psi_{\textrm{el}} (r, R)$ can seemingly be neglected and the Schr\"{o}dinger equation for 
$\phi_{\textrm{nucl}}(R)$ contains the information of electrons only through
the ground state energy $E(R)$ of electrons which depends
on the nuclear position $R$ regarded as the static parameter.
In this approximation, however, the nucleus feels the external 
electromagnetic field as the particle with positive charge $Ze$.
This drawback can be remedied by introducing the Berry 
phase into the Hamiltonian of nucleus \cite{mead1992geometric}.
\begin{align}
	H_{\textrm{nucl}} = \sum_{n} \frac{\left[\bm{P}_n - Ze\bm{A}_n -\bm{a}_n(R)\right]^2}{2m_{\textrm{nucl}}} + U(R), \label{eq:Hamiltonian}
\end{align}
where $n$ specifies the atom and $R=(\bm{R}_1, \cdots, \bm{R}_N)$
represents the coordinates of all the $N$ atoms. 
Here, $\bm{a}_n(R)$ is the Berry connection given by 
\begin{align}
	\bm{a}_n(R) = i\hbar \left\langle \psi_{\textrm{el}}(r,R) | \nabla_{\bm{R}_n} \psi_{\textrm{el}}(r,R) \right\rangle ,
  \label{eq:an}
\end{align}
where $ |\psi_{\textrm{el}}(R)\rangle  $ is the state of electrons with dependence 
on nuclear coordinates. $ U(R) $ is the sum of the electronic 
ground state energy and the interaction between nuclei.

This $\bm{a}_n(R)$ cancels the vector potential $\bm{A}_n$ 
for the external electromagnetic field in the case of single atom, 
i.e., the screening of the positive charge of nucleus by electrons 
is recovered~\cite{mead1992geometric}. 
For the hydrogen molecule, it has been discussed that 
this screening is perfect for the center-of-mass motion while
the magnetic field effect survives for the relative motion of the two 
atoms~\cite{ceresoli2007electron}. Therefore, the effect of the magnetic field on the
phonons in crystal remains an important issue to be studied.

In the present Letter, we study theoretically the 
Berry phase appearing in the phonon Hamiltonian and 
the consequent thermal Hall effect in a trivial band insulator.
Our model is the spinless fermion model with 1s orbital 
at each site, and there are no magnetic moments or 
spin-orbit interaction. Therefore, the effect of the 
magnetic field is only through the orbital motion of
electrons and nuclei. As for the electrons, the Lorentz force
is acting to produce the weak orbital diamagnetism but there is
no thermal Hall effect because of the energy gap in the low temperature
limit. As for the phonons, on the other hand, the acoustic phonons have
gapless dispersions, and hence can be excited thermally even at 
low temperature. 

\textit{Berry curvature and screening} --- 
From the Berry connection given in Eq.~(\ref{eq:an}), we obtain the
Berry curvature $F_{\mu\nu}$ as  
\begin{align}
F_{\mu\nu} =& \partial_\mu a_\nu - \partial_\nu a_\mu \notag\\
=& -2\hbar \textrm{Im} \left\langle 
\partial_\mu\psi_{\textrm{el}}(r,R)|\partial_\nu\psi_{\textrm{el}}(r,R)\right\rangle . \label{curv}
\end{align}
Here we introduced the symbol $ \mu=(n,\alpha) $ 
for the $ \alpha $-component  of $ n $-th nucleus.
Therefore, $F_{\mu \nu}$ is the tensor with $3N \times 3N$ components. 
We also denote it by $ F_{nm}^{\alpha\beta} $ in order to emphasize 
the nuclear and spacial indices. 
We note that, as mentioned by Resta~\cite{resta2000manifestations}, 
$ F_{\mu\nu} $ is antisymmetric for the exchange of $ \mu$ and $\nu $, 
but not for $ \alpha $ and $ \beta $, i.e., the condition
\begin{align}
F_{nm}^{\alpha\beta}= -F_{nm}^{\beta\alpha} =F_{mn}^{\alpha\beta} \label{antisym}
\end{align}
is not always true.

The EoM of nuclei is obtained from Eq.~\eqref{eq:Hamiltonian}:
\begin{align*}
M\ddot{R}_{n}^\alpha = -\partial_{n\alpha}U + \epsilon^{\alpha\beta\gamma}V_n^\beta Ze B^\gamma - \sum_{m} V_m^\beta F_{nm}^{\alpha\beta}.
\end{align*}
When Eq.~\eqref{antisym} holds, one can define a vector $ b_{nm}^\alpha:=\frac{1}{2}\epsilon^{\alpha\beta\gamma} F_{nm}^{\beta\gamma} = \frac{1}{2}\epsilon^{\alpha\beta\gamma} F_{mn}^{\beta\gamma} $, by which 
the last term turns into $ \sum_m \epsilon^{\alpha\beta\gamma}V_m^\beta b_{nm}^\gamma $. This means that $ \bm{b}_{nm} $ works as an effective magnetic field 
in the system and induces effective Lorentz force.

In a Hydrogen-like atom under 
magnetic field, the Berry curvature contribution cancels 
the external magnetic field~\cite{mead1992geometric}. 
The key point is the $ U(1) $ phase attached to the wave function 
due to the magnetic field. The atomic orbital $ \varphi(\bm{r}-\bm{R}) $ 
acquires an extra phase under the magnetic field:

\begin{align}
\varphi(\bm{r}-\bm{R}) \to \varphi'(\bm{r},\bm{R})
=\varphi(\bm{r}-\bm{R})e^{\frac{ie}{\hbar} \bm{A}(\bm{R})\cdot \bm{r}}. 
\label{phase}
\end{align}
Here, we have fixed the gauge and used the symmetric gauge. 
Although the manner of attaching the phase is a subtle problem, 
it is known that, as for the symmetric gauge, Eq.~\eqref{phase} yields 
physically correct results up to $ B $-linear order. 
In calculating the curvature, the derivative with respect to the nuclear coordinates 
is modified by this additional phase, which extracts the effect from the magnetic field.

On the other hand, the situation is quite different if the system contains two or more nuclei. 
In Hydrogen molecule, for instance, cancellation of the external magnetic field and 
the Berry curvature is perfect for the translational motion, 
but not for the relative motion of the two nuclei \cite{ceresoli2007electron}. In general cases, the screening of the magnetic field is guaranteed only for 
the translational motion, described by
\begin{align}
\sum_{nm}F_{nm}^{\alpha\beta}=-N\epsilon^{\alpha\beta\gamma}eB^\gamma.
\label{eq:screening_general}
\end{align}

\textit{Effective Hamiltonian for the band insulator} ---
As a theoretical model, we here study the Berry curvature of phonons in a two-dimensional 
square lattice with $ N $ nuclei and the same number of spinless electrons. 
Each electron is tightly bound to each atom, and then the wave 
function for the non-interacting electrons is given by the Slater determinant of the 
1s wave functions of all the $ N $ atoms. 
We denote the single-particle state of the $i$-th electron at the 
$n$th nucleus by $ \phi_{in}=\varphi'(r_i,R_n) $, given in Eq.~\eqref{phase}. 
The many-body wavefunction is proportional to the determinant of $ \Phi(r,R) $, 
an $ N\times N $ matrix whose $ (i,n) $-component is given by $ \phi_{in} $. 

The key factor which characterizes the Berry curvature of electron-phonon coupled systems is the overlap integral between the orbitals of $ n $-th and $ m $-th atoms, which we denote by $ S_{nm} $. The overlap integral is affected by the additional phase factor of the atomic orbital under the magnetic field. The modified overlap integral, $ S'_{nm}:=\int d^3r_i~\phi_{in}^* \phi_{im} $, is given by $ S_{nm}' = S_{nm}e^{i\theta_{nm}} + \mathcal{O}(B^2) $, where $ \theta_{nm}:=\frac{e}{\hbar}\bm{A}(\bm{R}_n)\cdot \bm{R}_m $. For brevity, we here define two matrices, $ S $ and $ S' $, whose $ (n,m) $ components are $ S_{nm} $ and $ S'_{nm} $, respectively.

The normalized wave function of this system is given by
\begin{align}
\psi_{\textrm{el}}(r,R) = (N!\det S')^{-\frac{1}{2}}\det \Phi(r,R).
\end{align} 
The general formula of $ F_{\mu\nu} $ of lattice systems is 
obtained by substituing $ \psi_{\textrm{el}}(r,R) $ into Eq.~\eqref{curv}, which is reduced to simpler 
formulae in concrete models, the details of which are given in the Supplemental Materials~\cite{suppl}. 
Here, we assume that the overlap integral between the 
nearest-neighbor nuclei is dominant, which is valid when the lattice constant is 
sufficiently large. 
Under this assumption, $F_{\mu\nu}$ for the square lattice is obtained up to the linear order in $B$~\cite{suppl}: 
\begin{subequations}
	\begin{align}
	a_n^\alpha =& \frac{-e}{\det S} 
	\Bigl[A_n^\alpha \widetilde{S}_{nn} + \sum_{l\neq n} A_l^\alpha \widetilde{S}_{nl} S_{nl}\Bigr],
	\label {conn_lattice} \\
	F_{nn}^{\alpha\beta} =& -\frac{\widetilde{S}_{nn}}{\det S} 
	\epsilon^{\alpha\beta\gamma}eB^\gamma 
	\notag\\&-\sum_{l}\frac{\widetilde{S}_{nl}}{\det S}
	\bigl[-e(A_n^\beta-A_l^\beta)\partial_{n\alpha}S_{nl} \notag\\
	&+ e(A_n^\alpha-A_l^\alpha)\partial_{n\beta}S_{nl}\bigr] ,
	\label{curv_lattice_a} \\
	F_{nm}^{\alpha\beta} =& -\frac{\widetilde{S}_{nm}}{\det S} \bigl[\epsilon^
	{\alpha\beta\gamma}eB^\gamma S_{nm} \notag\\
	&+ e(A_{n}^\beta-A_{m}^\beta)\partial_{n\alpha}S_{nm} \notag\\
	&- e(A_{n}^\alpha-A_{m}^\alpha)\partial_{m\beta}S_{nm}\bigr].
	\label{curv_lattice_b}
	\end{align}
\end{subequations}
Here, $ \widetilde{S}_{nm} $ is the $ (n,m) $-component of the cofactor matrix of $ S $, 
and we abbreviated $ A^\alpha(\bm{R}_n) $ to $ A_n^\alpha $. 
Equations~\eqref{curv_lattice_a}~and~\eqref{curv_lattice_b} imply that the curvature is mainly dependent on the overlap integral and its derivative.
The obtained expression of $ F_{nm}^{\alpha\beta} $ is antisymmetric 
for the exchange of $ \alpha $ and $ \beta $, and hence 
we can define a vector $ b_{nm}^\alpha:=\frac{1}{2}\epsilon^{\alpha\beta\gamma}F_{nm}^{\beta\gamma} $, the curvature felt by nucleus $ n $ and caused by nucleus $ m $. $ \bm{b}_{nm} $ is parallel to $ \bm{B} $, and the magnetic screening effect,
\begin{align}
\sum_m \bm{b}_{nm} = -Ze\bm{B},
\label{eq:screening_lattice}
\end{align}
is guaranteed, as in Eq.~\eqref{eq:screening_general}.
\begin{figure}
	\centering
	\includegraphics[bb=0 0 404 277,scale=0.6]{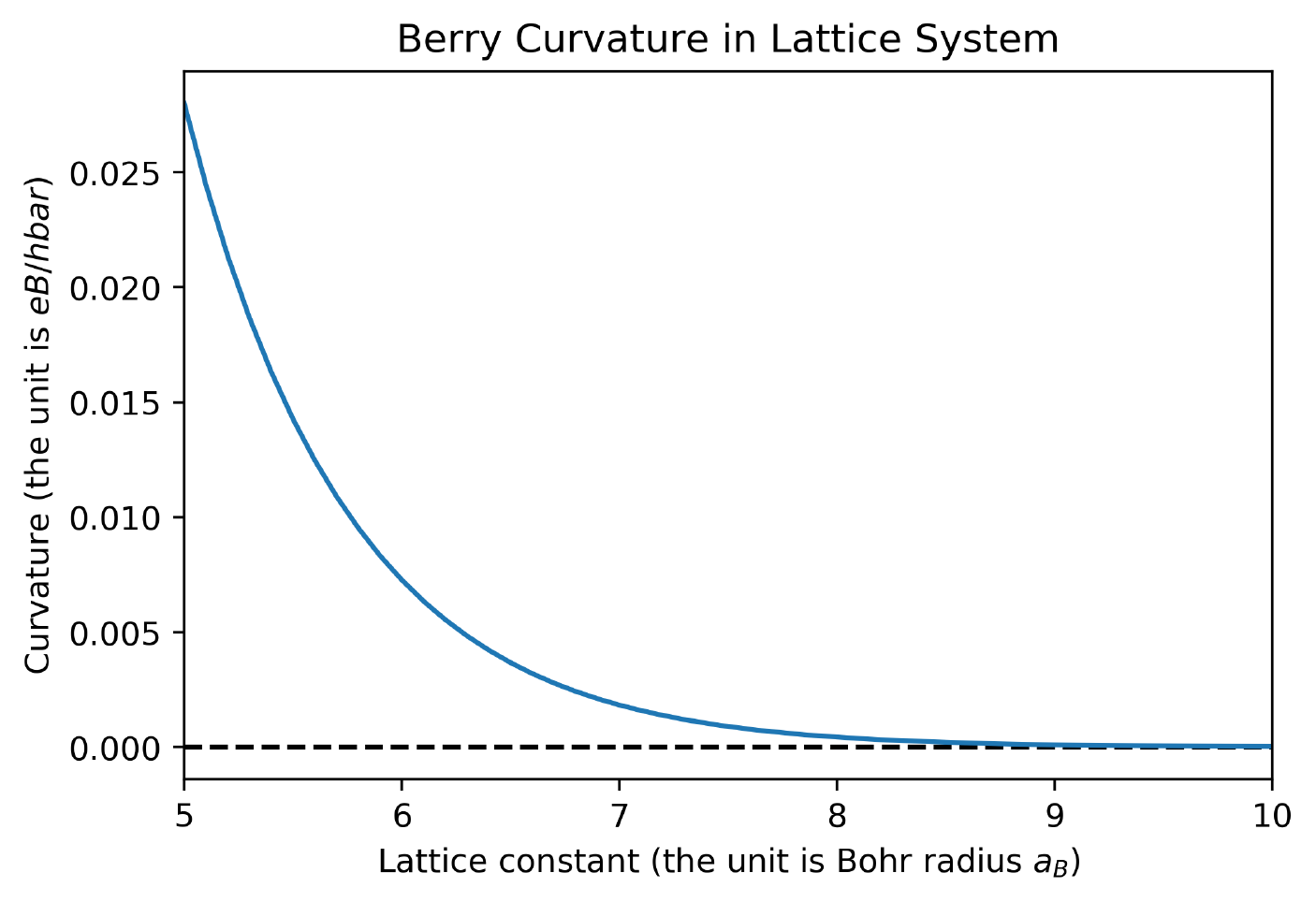}
	\caption{Molecular Berry curvature of the nearest-neighboring nuclei, $ F_{nm}^{xy} $ (or, equivalently, $ b^z_{nm} $), is plotted with respect to the lattice constant. The unit of the vertical and holozontal axis is taken as $ ZeB $ and the Bohr radius $ a_B $, respectively. The number of sites is $ N=20\times 20 $.}
	\label{lattice_curv}
\end{figure}

The effective Hamiltonian for the phonons are obtained by considering a small deviation $\bm{u}_n$ of the nuclei from its ground state position $\bm{R}_n^0$; $ \bm{R}_n = \bm{u}_n+\bm{R}_n^0 $. As for the interaction among nuclei, only quadratic terms of $ u_n^\alpha $'s are taken into account. The Berry curvature at the equilibrium atomic positions, i.e., $ \left. F_{nm}^{\alpha\beta}\right|_{R=R^0} $, is included in the Hamiltonian by the mininal coupling, as in Eq.~\eqref{eq:Hamiltonian}. Hereafter, we omit `$ R=R^0 $' and always take this substitution without notice. 
Figure \ref{lattice_curv} shows the numerical evaluation of $F_{nm}^{xy}$ 
for the nearest-neighbor pair $ (n,m) $. 
a
The Berry connection can be expressed by using $ \bm{b}_{nm} $ up to the linear order of $ u_n^\alpha $:
\begin{align}
\bm{a}_n=&\sum_m \frac{1}{2}\bm{b}_{nm}\times\bm{R}_m \notag\\
=& \sum_m \frac{1}{2}\bm{b}_{nm}\times\bm{u}_m+(\textrm{const.}),
\label{eq:vecpot1}
\end{align}
where the constant term can be removed by gauge transformation. 
Combining the contributions from the external magnetic field and the Berry curvature,
 the effective vector potential becomes
\begin{align}
\bm{a}_n' :=& \frac{Ze}{c} \bm{A}(\bm{R}_n) + \bm{a}_n \notag\\
=& \frac{1}{2}\sum_m \bm{b}_{nm} \times (\bm{u}_m-\bm{u}_n). \label{vecpot1}
\end{align}
In the second line, we used the identity Eq.~\eqref{eq:screening_lattice}. 

The Hamiltonian of the lattice vibration is modified in the exsistence of $ \bm{a}'_n $: 
the resultant Hamiltonian for our model is given by
\begin{subequations}
	\begin{align}
	H_{\textrm{nucl}}=&\sum_n \Bigl[ \frac{(\bm{P}_n-\bm{a}'_n)^2}{2m_{\textrm{nucl}}} + \frac{1}{2}
	\sum_m \bm{u}_n^TD_{nm}\bm{u}_m\Bigr] \\
	=& \sum_{\bm{k}} \left[\frac{1}{2m_{\textrm{nucl}}} \Pi_{\bm{k}}^{\alpha\dagger} \Pi_{\bm{k}}^\alpha 
	+ \frac{1}{2} D_{\bm{k}}^{\alpha\beta} u_{\bm{k}}^{\alpha\dagger} u_{\bm{k}}^{\beta}\right],
	\label{eq:Hamiltonian_phonon_k}
	\end{align}
\end{subequations}
where $ \Pi_{\bm{k}}:=P_{\bm{k}}-a'_{\bm{k}} $ is the momentum under a magnetic field. 
It is noted that the commutation relationship of $u_{\bm{k}}$ and $ \Pi_{\bm{k}} $ is different 
from that of usual canonical operators: $ [u_{\bm{k}}^\alpha, u_{\bm{q}}^\beta] = 0 $, 
$ [u_{\bm{k}}^\alpha,\Pi_{\bm{q}}^\beta] = i\hbar $, and $ [\Pi_{\bm{k}}^\alpha,
\Pi_{\bm{q}}^\beta] = i\hbar G_{\bm{k}}^{\alpha\beta} $, where 
$ G_{\bm{k}}^{\alpha\beta}:=\partial_{k\alpha}{a'}_{\bm{k}}^{\beta}-
\partial_{k\beta}{a'}_{\bm{k}}^{\alpha} $. 
We can see that the effective vector potential plays a role of the Raman interaction. 
Compared with the original Raman interaction, $ \bm{a}'_n $ does not include 
$ k $-independent constant terms but second-order derivative, so that it vanishes 
in the $ k\to 0 $ limit.
This is consistent with Eq.~(45) of Ref.~\onlinecite{qin2012berry}.

\textit{Thermal Hall effect} --- 
The effective vector potential $ \bm{a}'_n $ induces the thermal Hall effect of phonons.
In this section, we derive the analytic expression for the thermal Hall conductance, $ \kappa_H $, 
and numerically estimate it.

The definition of energy current in lattice systems, which we denote $ \bm{J}_E $, 
was given by Hardy~\cite{hardy1963energy} as an operator satisfying local energy conservation law. 
In addition to the usual Kubo formula, we have to take into account the contribution from the 
energy magnetization, $ \bm{M}_E $, defined by $ \bm{J}_E = \nabla \times \bm{M}_E $
~\cite{qin2011energy}.
The complete formula for the themal Hall conductance is given by
\begin{align*}
\kappa_H^\textrm{tr} = \kappa_H^\textrm{Kubo} + \frac{2M_E^z}{TV}.
\end{align*}
The detailed calculation was given by Qin {\it et al.}~\cite{qin2012berry}. 
General formula for the thermal Hall conductance of bosonic particles are:
\begin{align}
\kappa_H^\textrm{tr} = -\frac{(\pi k_B)^2T}{3\hbar} Z_\textrm{ph} - \frac{1}{T} \int_0^\infty d\epsilon^2~\sigma_{xy}(\epsilon)\frac{dn_B(\epsilon)}{d\epsilon},
\label{eq:condactance_general}
\end{align}
where
\begin{align}
Z_\textrm{ph}:=&\sum_{i\in\text{particle bands}}\frac{1}{V}\sum_{\bm{k}\in BZ} \Omega_{\bm{k}i}^z,\\
\sigma_{xy}(\epsilon) :=& -\frac{1}{V\hbar} \sum_{\hbar \omega_{\bm{k}i}\leq \epsilon} \Omega_{\bm{k}i}^z.
\end{align}
Here, $ \omega_{\bm{k}i} $ is the frequency of the phonon of $ i $-th mode and $ n_B(\epsilon) =1/(e^{\beta \epsilon}+1)$ is the Bose distribution.
The Berry curvature of phonons is defined as following~\cite{zhang2010topological,qin2012berry}: we define two matrices $ \mathcal{A}_{\bm{k}} $ and $ \mathcal{B}_{\bm{k}} $ by $ \mathcal{A}_k=\begin{bmatrix} O & i\hbar \\  -i\hbar & i\hbar G_{\bm{k}} \end{bmatrix} $ and $ \mathcal{B}_k = \begin{bmatrix} D_{\bm{k}} & O \\  O & 1/M\end{bmatrix} $. Here, $ \mathcal{A}_{\bm{k}} $ is the commutation relationship of $ \xi=\begin{bmatrix} \bm{u}_{\bm{k}} \\ \Pi_{\bm{k}} \end{bmatrix} $, and $ \mathcal{B}_{\bm{k}} $ is the bilnear form of the Hamiltonian in $ \xi $-basis. From the EoM, the eigenenergy is obtained by diagonalizing $ \tilde{H}_{\bm{k}}:=\mathcal{A}_{\bm{k}} \mathcal{B}_{\bm{k}} $. We denote the eigenvectors of $ \tilde{H}_{\bm{k}} $ by $ |v_i\rangle $. Then the Berry connection and curvature are defined by
\begin{subequations}
	\begin{align}
	a_{\bm{k}i}^\alpha :=& -\textrm{Im}~ \langle v_i| \mathcal{B}_k\partial_{k\alpha}|v_i\rangle, \\
	\Omega_{\bm{k},i}^z :=& \partial_{kx} a_{\bm{k}i}^y - \partial_{ky} a_{\bm{k}i}^x.
	\end{align}
\end{subequations}

The second term in Eq.~\eqref{eq:condactance_general} consists of a summation of Berry curvature 
over the Brillouin zone and over all the particle bands, $ Z_\textrm{ph} $. 
In the previous study, $ Z_\textrm{ph} $ is supposed to vanish in most cases \cite{qin2012berry}. 
For the case of magnonic systems, the 
summation of Chern number over particle bands is exactly zero~\cite{shindou2013topological}. This can be explained 
by the fact that the BdG Hamiltonian of the system is adiabatically connected to a trivial matrix, 
whose Berry curvature is zero. A parallel discussion leads to the conclusion that 
$ Z_\textrm{ph}=0 $ holds exactly in our model~\cite{suppl}, if we perturb the system to introduce the gap at $ \bm{k}=0 $ so that the Chern number is well-defined. 

On the estimation of $ \kappa_H $, it is convenient to introduce the continuum approximation. 
The dynamical matrix and the vector potential of the general Hamiltonian of phonons with Raman-type interaction are given by 
$ D_{\bm{k}}^{\alpha\beta}=\mu_1 k^2 \delta^{\alpha\beta} +\mu_2 k^\alpha k^\beta $ and 
$ {a'}_{\bm{k}}^\alpha = \gamma_1\partial_\alpha \partial_\beta \epsilon^{\beta\rho\sigma} 
M^\rho u^\sigma + \gamma_2 \nabla^2 \epsilon^{\alpha\beta\gamma}M^\beta u^\gamma $ in Eq.~\eqref{eq:Hamiltonian_phonon_k}, where $ \gamma_1 $ and $ \gamma_2 $ are coupling constants \cite{qin2012berry}. The corresponding geometric curvature is $ G_{\bm{k}}^{\alpha\beta} = \frac{1}{m_{\textrm{nucl}}} 
\epsilon^{\alpha\beta\gamma} (-\gamma_1k^\gamma \bm{k}\cdot\bm{M}+
(\gamma_1+2\gamma_2)k^2M^\gamma) $.
As long as the temperature is sufficiently low, i.e., the condition $ k_BT \ll \hbar\omega_{D,i} $ 
is satisfied for each band $ i $ ($ \omega_{D,i}:=c_i\pi/a $ is the Debye frequency of $ i $-th band), 
the contribution from small $ k $ is donimant since the derivative of the Bose distribution function, 
$ dn_B(\epsilon)/d\epsilon $, decreases drastically as $ \hbar \omega_{\bm{k}i} $ 
becomes higher. 
Therefore, at sufficiently low temperature, $ \kappa_H $ is well estimated by assuming that the 
continuum approximation is valid over the whole Brillouin zone. Under these conditions, the thermal Hall conductance in two dimension can be calculated by

\begin{align}
\kappa_H^\textrm{2D} =& -\frac{(\pi k_B)^2T}{3\hbar}Z_\textrm{ph} + \frac{\Gamma^\textrm{2D} k_B^3T^2}{m_{\textrm{nucl}}\hbar^2c_T^2} \int_{0}^\infty dx~\frac{x^3e^x}{(e^x-1)^2}. \label{eq:conductance2D}
\end{align}
\begin{figure}
	\centering
	\includegraphics[bb=0 0 366 227,scale=0.6]{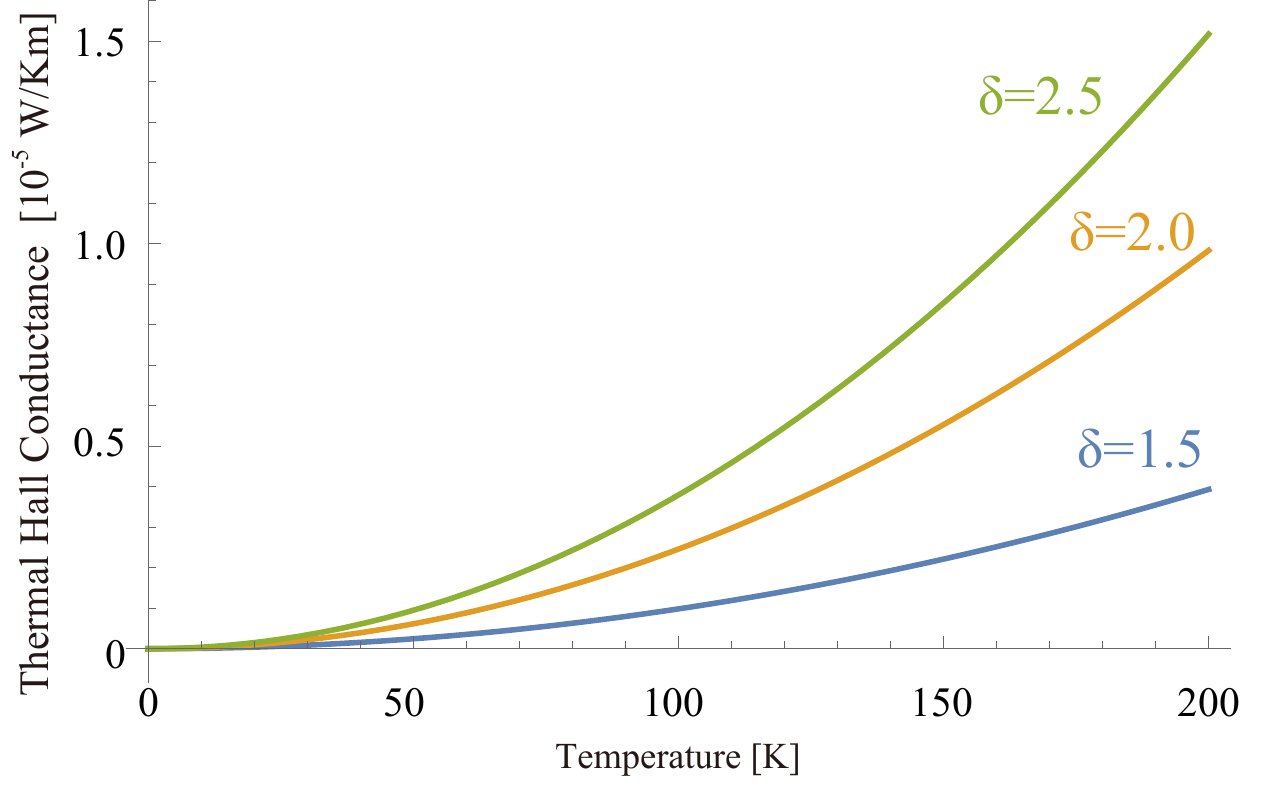}
	\caption{The thermal Hall conductance is plotted with respect to temperature. The blue,  orange and green lines correspond to $ \delta=$1.5, 2.0  and 2.5, respectively. The used parameters are: $ a=6.5a_B $, $ c_T=3000 $[m/s] and $ B=10 $[T].}
	\label{fig:condctance}
\end{figure}
$ \Gamma^\textrm{2D} $ is a constant dependent on the ratio of sound speed of the longitudinal 
and transverse modes, $ \delta:=c_L/c_T $, which is given by 
\begin{align}
\Gamma^\textrm{2D} =& 2\pi(2\gamma_2-\gamma_1)M^z\frac{(\delta-1)^2}{\delta(\delta+1)}.
\end{align}

Now, we apply the discussion above to our model. In nonmagnetic insulators, the geometric curvature, $ \bm{b}_{nm} $, plays the role of the magnetization $ \bm{M} $, and it takes finite value only for nearest-neighbor nuclei. 
We here denote $ \bm{b}_{nm} $ of nearest-neighbor nuclei by $ \bm{b} $. 
The sum over $ m $ in Eq.~\eqref{vecpot1} is reduced to the sum over nearest-neighbor 
nuclei, i.e, $m$ is an integer which satisfies 
$ \bm R_m^0=\bm R_n^0\pm \bm e_x, \bm R_n^0\pm \bm e_y $ where 
$\bm e_x $ ($\bm e_y $) is the lattice vector for $x$ ($y$)-direction. Assuming $ \bm{u}_n $ is a slowly-varying parameter of $\bm R_n$, we expand it by the gradient as 
$ u_{n\pm x(y)}^\alpha \simeq u_n^\alpha \pm a \partial_{x(y)} u_n^\alpha + \frac{a^2}{2} \partial_{x(y)}^2 u_n^\alpha $. 
Then the vector potential \eqref{vecpot1} becomes
\begin{align}
\bm{a}'(\bm{R})= a^2\bm{b}\times\nabla^2\bm{u}(\bm{R}). 
\end{align}
Therefore the coupling constants are repleced as $ \gamma_1\to0 $ and $ \gamma_2M^z\to a^2b^z $. 

The dependence on $ T $ of $ \kappa_H^\textrm{2D} $ 
is shown in Figure \ref{fig:condctance} by using typical parameters. 
Here, the thermal Hall conductance in two-dimensions 
is translated into one in three-dimensions by assuming that the thickness of the layer is almost 
the same as the lattice constant, i.e., 
the plotted value is 
$ \tilde{\kappa}_H:=\kappa_H^\textrm{2D}/a $. Experimentally, the thermal Hall conductance 
in magnetic materials is $ \kappa_H \simeq 10^{-5} $[W/Km] at $ T\simeq 5 $[K] 
\cite{strohm2005phenomenological, inyushkin2007phonon}. 
Compared to this, $ \kappa_H $ of our model is much smaller at 
the same temperature. However, it is detectable enough at higher temperatures. 
For the experimental realization, Eq.~\eqref{eq:conductance2D} implies that materials with larger overlap integral and smaller transverse sound speed are expected to show larger 
thermal Hall conductance even without magnetization.
\\

This work was supported by JSPS KAKENHI Grant Nos. JP18H03676, JP18H04222, JP24224009, and JP26103006, ImPACT Program of Council for Science, Technology and Innovation (Cabinet office, Government of Japan), and CREST, JST (Grant No. JPMJCR1874). T.S. was supported by Japan Society for the Promotion of Science through Program for Leading Graduate Schools (MERIT).

\end{document}


\title{Supplemental Materials: \textit{Berry phase of phonons and thermal Hall effect in nonmagnetic insulators}}
	\maketitle
		
\newpage
\section{I. Molecular Berry curvature of lattice systems}
We here show how Eqs.~(10a),~(10b)~and~(10c) in the main text are derived. 
Using the wave function given by the Slater determinant of all the atoms, 
$ \psi_{\textrm{el}}(r,R) = (N!\det S')^{-\frac{1}{2}}\det \Phi(r,R) $, 
the Berry connection and curvature is calculated as
\begin{align}
a_n^\alpha =& i\hbar \langle \psi_{\textrm{el}} |\partial_{n\alpha} \psi_{\textrm{el}} \rangle \notag\\
=& i\hbar \bigl(\mathcal{N} \partial_{n\alpha} \mathcal{N} |\Phi|^*|\Phi| 
+ \mathcal{N}^2 |\Phi|^* \partial_{n\alpha} |\Phi|\bigr)
\label{eq:BerryConn} 
\end{align}
\begin{align}
F_{nm}^{\alpha\beta} 
=& \partial_{n\alpha} a_m^\beta - \partial_{m\beta} a_n^\alpha 
\notag\\
=& -2\hbar~{\rm Im} \left\langle \partial_{n\alpha} \psi_{\textrm{el}}|\partial_{m\beta}\psi_{\textrm{el}}\right\rangle \notag\\
=& -2\hbar~{\rm Im} \int d^3r_1\cdots d^3r_N~\partial_{n\alpha}\psi_{\textrm{el}}^* \partial_{m\beta} \psi_{\textrm{el}} \notag\\
=& -2\hbar~{\rm Im}\Bigl[(\partial_{n\alpha} \mathcal{N})(\partial_{m\beta} \mathcal{N}) \int d^{3N}r~ |\Phi|^2 \notag\\
&+ \mathcal{N} (\partial_{n\alpha} \mathcal{N}) \int d^{3N}r~|\Phi|^* \partial_{m\beta} |\Phi|  \nonumber\\&+ \mathcal{N} (\partial_{m\beta} \mathcal{N}) \int d^{3N}r~|\Phi| \partial_{n\alpha} |\Phi|^* \notag\\
&+ \mathcal{N}^2 \int d^{3N}r~ \partial_{n\alpha} |\Phi|^* \partial_{m\beta}|\Phi| \Bigr].
\label{eq:BerryCurv}
\end{align}

In the following, some formulae are provided so as to simplify the analytic form of Eq.~\eqref{eq:BerryCurv}. Since we are interested in the dependence on the magnetic field, $ B $, we only extract $ B $-linear terms. 		
		
		\subsection{Some formulae for the calculation of $ F_{\mu\nu} $}

		\subsubsection{The properties of $ S' $}
		The derivative of $ S' $ is given by
		\begin{subequations}
			\begin{align}
			\partial_{n\alpha} S'_{nm} =& e^{i\theta_{nm}} [\partial_{n\alpha}S_{nm} - ieA_m^\alpha S_{nm}] \\
			\partial_{m\beta} S'_{nm} =& e^{i\theta_{nm}} [\partial_{m\beta}S_{nm} + ieA_n^\beta S_{nm}].
			\end{align}			
		\end{subequations}
		We here note that, since $ S_{nm} $ is dependent on the distance between $ n $ and $ m $, $ S_{nm}=S_{mn} $ and $ \partial_{n\alpha}S_{nm} = -\partial_{m\alpha}S_{nm} $ hold.
		
		Since $ S' $ is Hermitian by definition, $ \det S' $ is real-valued. In addition, since the wavefunction $\phi$ is real, the expansion of $ \det S' $ in the imaginary unit $i$ is equivalent to the expansion in $B$. Therefore, there is no $ B $-linear term in $ \det S' $:
		\begin{align}
		\det S' = \det S + \mathcal{O}(B^2).
		\end{align}
		The determinant of a matrix $ X $ is related with its inverse matrix $ X^{-1} $ and cofactor matrix $ \widetilde{X} $ as
		\begin{align}
		X^{-1} = \frac{\widetilde{X}}{\det X}.
		\end{align}
		From here on, we denote a cofactor matrix by $ \widetilde{~~~} $, and the $ (k,l) $ cofactor of the $ (n,m) $-submatrix of $ X $ by $ \widetilde{X}_{mn;lk} $, where an $ (n,m) $-submatrix is defined as the $ (N-1)\times (N-1) $ matrix $ Y^{(n,m)} $, given by
		\begin{align}
		&Y^{(n,m)} \notag\\
		:=& \left[\begin{array}{cccccc}
		X_{1,1} & \cdots & X_{1,m-1} & X_{1,m+1} & \cdots & X_{1,N} \\
		\vdots &\ddots&\vdots&\vdots&\ddots&\vdots\\
		X_{n-1,1} &\cdots&X_{n-1,m-1}&X_{n-1,m+1}&\cdots&X_{n-1,N}\\
		X_{n+1,1} &\cdots&X_{n+1,m-1}&X_{n+1,m+1}&\cdots&X_{n+1,N}\\
		\vdots &\ddots&\vdots&\vdots&\ddots&\vdots\\
		X_{N,1} &\cdots&X_{N,m-1}&X_{N,m+1}&\cdots& X_{N,N}
		\end{array}\right].
		\end{align}
		We note that the correspondence between the indices of $ X $ ($ i $ and $ j $) and those of $ Y^{(n,m)} $ ($ \bar{i}$ and $\bar{j} $) is given by one-to-one mappings $ \tau_n:~\{1,\cdots,N\}\setminus \{n\}\to \{1,\cdots,N-1\} $ and $ \tau_m:~\{1,\cdots,N\}\setminus\{m\}\to \{1,\cdots,N-1\} $ such that
		\begin{subequations}
			\begin{align}
			\bar{i}=&\tau_n(i) = \begin{cases}
			i & i<n\\
			i-1 & i>n
			\end{cases}, \\
			\bar{j}=&\tau_m(j) = \begin{cases}
			j & j<m\\
			j-1 & j>m
			\end{cases}.
			\end{align}
		\end{subequations}
		The $ (\bar{k},\bar{l}) $-submatrix of $ Y^{(n,m)} $ is the $ (N-2)\times(N-2) $ matrix $ Z^{(\bar{k},\bar{l})} $, given by
		\begin{align}
		&Z^{(\bar{k},\bar{l})} \notag\\
		:=&
		\left[\begin{array}{cccccc}
		Y^{(n,m)}_{1,1} & \cdots & Y^{(n,m)}_{1,\bar{l}-1} & Y^{(n,m)}_{1,\bar{l}+1} & \cdots & Y^{(n,m)}_{1,N-1} \\
		\vdots &\ddots&\vdots&\vdots&\ddots&\vdots\\
		Y^{(n,m)}_{\bar{k}-1,1} &\cdots&Y^{(n,m)}_{\bar{k}-1,\bar{l}-1}&Y^{(n,m)}_{\bar{k}-1,\bar{l}+1}&\cdots&Y^{(n,m)}_{\bar{k}-1,N-1}\\
		Y^{(n,m)}_{\bar{k}+1,1} &\cdots&Y^{(n,m)}_{\bar{k}+1,\bar{l}-1}&Y^{(n,m)}_{\bar{k}+1,\bar{l}+1}&\cdots&Y^{(n,m)}_{\bar{k}+1,N-1}\\
		\vdots &\ddots&\vdots&\vdots&\ddots&\vdots\\
		Y^{(n,m)}_{N-1,1} &\cdots&Y^{(n,m)}_{N-1,\bar{l}-1}&Y^{(n,m)}_{N-1,\bar{l}+1}&\cdots& Y^{(n,m)}_{N-1,N-1}
		\end{array}\right].
		\end{align}
		Then the two kinds of cofactor matrices are given by
		\begin{widetext}
			\begin{align}
			\widetilde{X}_{ji} 
			= (-1)^{i+j}\det Y^{(n,m)} 
			= (-1)^{i+j}\left|\begin{array}{cccccc}
			X_{1,1} & \cdots & X_{1,j-1} & X_{1,j+1} & \cdots & X_{1,N} \\
			\vdots &\ddots&\vdots&\vdots&\ddots&\vdots\\
			X_{i-1,1} &\cdots&X_{i-1,j-1}&X_{i-1,j+1}&\cdots&X_{i-1,N}\\
			X_{i+1,1} &\cdots&X_{i+1,j-1}&X_{i+1,j+1}&\cdots&X_{i+1,N}\\
			\vdots &\ddots&\vdots&\vdots&\ddots&\vdots\\
			X_{N,1} &\cdots&X_{N,j-1}&X_{N,j+1}&\cdots& X_{N,N}
			\end{array}\right|,
			\end{align}
			\begin{align}
			\widetilde{X}_{ji;lk} 
			=& (-1)^{\bar{k}+\bar{l}} \det Z^{(\bar{k},\bar{l})} \notag\\
			=& (-1)^{\bar{k}+\bar{l}}\left|\begin{array}{cccccc}
			Y^{(n,m)}_{1,1} & \cdots & Y^{(n,m)}_{1,\bar{l}-1} & Y^{(n,m)}_{1,\bar{l}+1} & \cdots & Y^{(n,m)}_{1,N-1} \\
			\vdots &\ddots&\vdots&\vdots&\ddots&\vdots\\
			Y^{(n,m)}_{\bar{k}-1,1} &\cdots&Y^{(n,m)}_{\bar{k}-1,\bar{l}-1}&Y^{(n,m)}_{\bar{k}-1,\bar{l}+1}&\cdots&Y^{(n,m)}_{\bar{k}-1,N-1}\\
			Y^{(n,m)}_{\bar{k}+1,1} &\cdots&Y^{(n,m)}_{\bar{k}+1,\bar{l}-1}&Y^{(n,m)}_{\bar{k}+1,\bar{l}+1}&\cdots&Y^{(n,m)}_{\bar{k}+1,N-1}\\
			\vdots &\ddots&\vdots&\vdots&\ddots&\vdots\\
			Y^{(n,m)}_{N-1,1} &\cdots&Y^{(n,m)}_{N-1,\bar{l}-1}&Y^{(n,m)}_{N-1,\bar{l}+1}&\cdots& Y^{(n,m)}_{N-1,N-1}
			\end{array}\right| \notag\\
			=& (-1)^{i+j+k+l} \left|\begin{array}{ccccccccc}
			X_{1,1} & \cdots & X_{1,j-1} & X_{1,j+1} & \cdots & X_{1,l-1} & X_{1,l+1} &\cdots & X_{1,N}\\
			\vdots &\ddots&\vdots&\vdots&\ddots&\vdots&\vdots&\ddots&\vdots\\
			X_{i-1,1} &\cdots&X_{i-1,j-1}&X_{i-1,j+1}&\cdots&X_{i-1,l-1}&X_{i-1,l+1}&\cdots&X_{i-1,N}\\
			X_{i+1,1}&\cdots&X_{i+1,j-1}&X_{i+1,j+1}&\cdots&X_{i+1,l-1}&X_{i+1,l+1}&\cdots&X_{i+1,N}\\
			\vdots &\ddots&\vdots&\vdots&\ddots&\vdots&\vdots&\ddots&\vdots\\
			X_{k-1,1}&\cdots&X_{k-1,j-1}&X_{k-1,j+1}&\cdots&X_{k-1,l-1}&X_{k-1,l+1}&\cdots&X_{k-1,N}\\
			X_{k+1,1} &\cdots&X_{k+1,j-1}&X_{k+1,j+1}&\cdots&X_{k+1,l-1}&X_{k+1,l+1}&\cdots&X_{k+1,N}\\
			\vdots &\ddots&\vdots&\vdots&\ddots&\vdots&\vdots&\ddots&\vdots\\
			X_{N,1} &\cdots&X_{N,j-1}&X_{N,j+1}&\cdots&X_{N,l-1}&X_{N,l+1}&\cdots&X_{N,N}
			\end{array}\right| .
			\end{align} 
		\end{widetext}
		In the last line, we used the fact that both acting $ \tau_n $ on $ i $ (or $ \tau_m $ on $ j $) and the exchange of $ i $ and $ k $ (or $ j $ and $ l $) yield the same factor $ -1 $ and therefore the ingredient of the determinant is always expressed in the order shown above. In the same way as $ \det S' $, $ B $-linear terms of $ \tilde{S'}_{nm} $ and $ \tilde{S'}_{nm;kl} $ are zero.
		
		\subsubsection{$ \mathcal{N} $ and $ \partial_\mu \mathcal{N} $}
		\begin{align}
		\mathcal{N} =& \frac{1}{\sqrt{N!\det S'}} = \frac{1}{\sqrt{N!\det S}} + \mathcal{O}(B^2) \\
		\partial_\mu \mathcal{N} =&-\frac{1}{2}\frac{1}{\sqrt{N!}}\frac{\partial_\mu\det S'}{(\det S')^{3/2}} \nonumber\\
		=&-\frac{1}{2}\frac{1}{\sqrt{N!}}\frac{\partial_\mu\det S}{(\det S)^{3/2}}+\mathcal{O}(B^2) \nonumber\\
		=& -\frac{1}{2} \frac{1}{\sqrt{N!}} \frac{{\rm tr} [S^{-1}\partial_\mu S]}{(\det S)^{1/2}} + \mathcal{O}(B^2) \notag\\
		=& -\frac{1}{\sqrt{N!}(\det S)^{3/2}} \sum_{l(\neq n)} \widetilde{S}_{nl}\partial_{n\alpha}S_{nl}+\mathcal{O}(B^2).
		\label{partial_N}
		\end{align}
		From the 2nd line to the 3rd line of \eqref{partial_N}, the formula $ \partial_\mu \det S=\det S\times{\rm tr}[S^{-1}\partial_\mu S] $ is used.
		In the last line, since the derivative $ \partial_{n\alpha} $ only hits the $ n $-th row and column of $ S $, 
		\begin{align}
		{\rm tr} [S^{-1} \partial_{n\alpha} S] 
		=& \sum_{l(\neq n)}\left[S^{-1}_{nl} \partial_{n\alpha} S_{ln} + S^{-1}_{ln} \partial_{n\alpha} S_{nl}\right ] \notag\\
		=& 2\sum_{l(\neq n)} S^{-1}_{nl} \partial_{n\alpha} S_{nl}\notag\\
		=& \frac{2}{\det S} \sum_{l(\neq n)} \widetilde{S}_{nl}\partial_{n\alpha}S_{ln},
		\label{trace_S}
		\end{align}
		In the 2nd line of \eqref{trace_S}, $ S_{nl}=S_{ln} $ and $ S^{-1}_{nl}=S^{-1}_{ln} $ are used. Due to $ S_{nn}=1 $, the summation over $ l $ need not include $ l=n $.
		\begin{align}
		&\mathcal{N}\partial_{n\alpha} \mathcal{N} \notag\\
		=& -\frac{{\rm tr} [S^{-1}\partial_\mu S]}{2N!\det S} +\mathcal{O}(B^2) \notag\\
		=& -\frac{1}{N! (\det S)^2} \sum_{l\neq n} \widetilde{S}_{nl}\partial_{n\alpha} S_{ln} + \mathcal{O}(B^2) .
		\end{align}
		\begin{align}
		&\partial_{n\alpha} \mathcal{N} \times \partial_{m\beta} \mathcal{N} \notag\\
		=& \frac{1}{N!(\det S)^3} \sum_{\substack{l\neq n\\l'\neq m}}[\widetilde{S}_{nl} \partial_{n\alpha} S_{ln}][\widetilde{S}_{ml'} \partial_{m\beta} S_{l'm}] + \mathcal{O}(B^2)
		\end{align}
		\begin{align}
		&\int d^{3N}r~ \det\Phi^*\det\Phi  \notag\\
		=& N!\det S' = N!\det S +\mathcal{O}(B^2) \notag\\
		&\int d^{3(N-1)}r_i~ \widetilde{\Phi}_{ni}^*\widetilde{\Phi}_{mi} \notag\\
		=& (N-1)! \widetilde{S'}_{mn} = (N-1)!\widetilde{S}_{mn}+\mathcal{O}(B^2) \notag\\
		&\int d^{3(N-2)}r_i~ \widetilde{\Phi}_{ni;mj}^*\widetilde{\Phi}_{ki;lj}  \notag\\
		=& (N-2)! \widetilde{S'}_{kn;lm} = (N-2)! \widetilde{S}_{kn;lm}+\mathcal{O}(B^2)
		\end{align}
		
		\subsubsection{The Berry connection and curvature of atomic orbitals}
		\begin{align}
		&\langle\phi_n|\partial_{n\alpha}\phi_n\rangle=\int d^3r_i~ \phi_{i,n}^* \partial_{n\alpha}\phi_{i,n} \notag\\
		=&\int d^3r_i~ \varphi(\bm{r}_i-\bm{R}_n)e^{i\theta_{in}} \notag\\
		&\times[\partial_{n\alpha}\varphi(\bm{r}_i-\bm{R}_n)-ieA_\alpha(\bm{r}_i)\varphi(\bm{r}_i-\bm{R}_n)]e^{-i\theta_{in}} \notag\\
		=&\int d^3r_i~ [\varphi(\bm{r}_i-\bm{R}_n)\partial_{n\alpha}\varphi(\bm{r}_i-\bm{R}_n) \notag\\
		&-ie(A_\alpha(\bm{r}_i-\bm{R}_n)+A_\alpha(\bm{R}_n)) \varphi(\bm{r}_i-\bm{R}_n)^2]\notag\\
		=& -ieA_\alpha (\bm{R}_n)
		\end{align}
		In the 3rd line, we used $ \int d^3r_i~A_\alpha(\bm{r}_i-\bm{R}_n) \varphi(\bm{r}_i-\bm{R}_n)^2=0 $ due to odd parity of the integrand.
		\begin{align}
		\langle\phi_n|\partial_{n\alpha}\phi_m\rangle=&\int d^3r_i~ \phi_{i,n}^* \partial_{m\beta}\phi_{i,m} \notag\\
		=& \partial_{n\alpha} S'_{nm} \notag\\
		=& e^{i\theta_{nm}} [\partial_{n\alpha}S_{nm} - iA_m^\alpha S_{nm}].
		\end{align}
		
		\subsubsection{The integrals of $ |\Phi| $}
		Using these integrals, 
		\begin{align}
		&\int d^{3N}r~ |\Phi|^*\partial_{n\alpha}|\Phi| \nonumber\\
		=& \int d^{3N}r~ |\Phi|^* \sum_j \widetilde{\Phi}_{nj}\partial_{n\alpha}\phi_{jn} \nonumber\\
		=& \sum_j\sum_l [\int d^{3(N-1)}r~ \widetilde{\Phi}_{lj}^*\widetilde{\Phi}_{nj}] [\int d^3r_j~\phi_{jl}^*\partial_{n\alpha}\phi_{jn}] \notag\\
		=& (N-1)! \sum_j[\widetilde{S'}_{nn}\lrangle{\phi_n|\partial_{n\alpha}\phi_n}+ \sum_{l\neq n} \widetilde{S'}_{nl} \partial_{n\alpha}S'_{ln}] \notag\\
		=& N! [\widetilde{S'}_{nn}(ieA_n^\alpha)+\sum_{l\neq n} \widetilde{S'}_{nl} \partial_{n\alpha}S'_{ln}] \notag\\
		=& N! [ieA_n^\alpha\widetilde{S}_{nn}+\sum_{l\neq n} \widetilde{S}_{nl} (\partial_{n\alpha}S_{nl} +ieA_{l}^\alpha  S_{nl}) ] \notag\\
		&+ \mathcal{O}(B^2) \notag\\
		=& N! [\sum_l ieA_{l}^\alpha  \widetilde{S}_{nl} S_{ln} + \sum_{l\neq n} \widetilde{S}_{nl} \partial_{n\alpha}S_{ln}] + \mathcal{O}(B^2)
		\end{align}
		\begin{align}
		&\int d^{3N}r~ \partial_{n\alpha}|\Phi|^*\partial_{n\beta}|\Phi| \nonumber\\
		=& \int d^{3N}r~ \sum_{i,j} [\widetilde{\Phi}_{ni}^* \partial_{n\alpha}\phi_{in}^*] [\widetilde{\Phi}_{nj} \partial_{n\beta}\phi_{jn}] \notag\\
		=& \sum_i\int d^{3(N-1)}r~ \widetilde{\Phi}_{ni}^*\widetilde{\Phi}_{ni} \int d^3r_i~ \partial_{n\alpha}\phi_{in}^* \partial_{n\beta}\phi_{in} \notag\\
		&+ \sum_{i\neq j}\sum_{l,l'\neq n} [\int d^{3(N-2)}r~\widetilde{\Phi}_{ni;lj}^*\widetilde{\Phi}_{nj;l'i} \notag\\
		&\times\int d^3r_i~\phi_{il'}\partial_{n\alpha}\phi_{in}^*\int d^3r_j~\phi_{jl}^*\partial_{n\beta}\phi_{jn}] \notag\\
		=& \sum_i (N-1)! \widetilde{S'}_{nn} \langle\partial_{n\alpha}\phi_{in}| \partial_{n\beta}\phi_{in}\rangle \notag\\
		&+ \sum_{i\neq j}\sum_{l,l'\neq n} (N-2)! \widetilde{S'}_{nn;l'l} \langle\partial_{n\alpha}\phi_{n}|\phi_{l'}\rangle\langle\phi_{l}|\partial_{n\beta}\phi_{n}\rangle \notag\\
		=& N! \widetilde{S}'_{nn} i\epsilon^{\alpha\beta\gamma}eB^\gamma/2 \notag\\
		&+ N! \sum_{l,l'\neq n} \widetilde{S}'_{nn;l'l} \partial_{n\alpha} S'_{nl'}\partial_{n\beta}S'_{ln} \notag\\
		=& N! \widetilde{S}_{nn} i/2 \notag\\
		&+ N!\sum_{l,l'\neq n} \widetilde{S}_{nn;l'l} e^{i(\theta_{nl'}-\theta_{ln})}\notag\\
		&\times [\partial_{n\alpha}S_{nl'} -ieA_{l'}^\alpha S_{nl'}][\partial_{n\beta}S_{ln} +ieA_{l}^\beta S_{ln}] +\mathcal{O}(B^2) \notag\\
		=& N! \widetilde{S}_{nn} i\epsilon^{\alpha\beta\gamma}eB^\gamma/2 \notag\\
		&+ N!\sum_{l,l'\neq n} \widetilde{S}_{nn;l'l} \times [e^{i(\theta_{nl'}-\theta_{ln})} \partial_{n\alpha}S_{nl'}\partial_{n\beta}S_{ln} \notag\\ 
		& -ieA_{l'}^\alpha S_{nl'}\partial_{n\beta}S_{ln} +ieA_{l}^\beta S_{ln}\partial_{n\alpha}S_{nl'}] +\mathcal{O}(B^2)
		\end{align}
		For $ n\neq m $,
		\begin{align}
		&\int d^{3N}r~ \partial_{n\alpha}|\Phi|^*\partial_{m\beta}|\Phi| \nonumber\\
		=&\int d^{3N}r~ \sum_{i,j} [\widetilde{\Phi}_{ni}^* \partial_{n\alpha}\phi_{in}^*] [\widetilde{\Phi}_{mj} \partial_{m\beta}\phi_{jm}] \notag\\
		=& \sum_i\int d^{3(N-1)}r~ \widetilde{\Phi}_{ni}^*\widetilde{\Phi}_{mi} \int d^3r_i~ \partial_{n\alpha}\phi_{in}^* \partial_{m\beta}\phi_{im} \notag\\
		&+ \sum_{i\neq j}\sum_{l\neq n}\sum_{l'\neq m} [\int d^{3(N-2)}r~\widetilde{\Phi}_{ni;lj}^*\widetilde{\Phi}_{mj;l'i} \notag\\
		&\times\int d^3r_i~\phi_{il'}\partial_{n\alpha}\phi_{in}^*\int d^3r_j~\phi_{jl}^*\partial_{m\beta}\phi_{jm}] \notag\\
		=& \sum_i (N-1)! \widetilde{S'}_{mn} \langle\partial_{n\alpha}\phi_{in}| \partial_{m\beta}\phi_{im}\rangle \notag\\
		&+ \sum_{i\neq j}\sum_{l\neq n}\sum_{l'\neq m} (N-2)! \widetilde{S'}_{mn;l'l} \langle\partial_{n\alpha}\phi_{n}|\phi_{l'}\rangle\langle\phi_{l}|\partial_{m\beta}\phi_{m}\rangle \notag\\
		=& N! \widetilde{S'}_{mn}  ie(\partial_{n\alpha}A_n^\beta S_{nm} + A_n^\beta\partial_{n\alpha}S_{nm} - A_m^\alpha\partial_{m\beta}S_{nm}) \notag\\
		&+ N!\sum_{l\neq n,m} \widetilde{S'}_{mn;nl} ieA_n^\alpha \partial_{m\beta}S'_{lm} \notag\\
		&+ N!\sum_{l'\neq n,m} \widetilde{S'}_{mn;l'm} (-ie)A_m^\beta \partial_{n\alpha}S^{'*}_{l'n} \notag\\
		&+ N!\widetilde{S'}_{mn;nm} ieA_n^\alpha (-ie)A_m^\beta \notag\\
		&+ N!\sum_{l,l'\neq n,m} \widetilde{S'}_{mn;l'l} \partial_{n\alpha} S_{l'n}^{'*} \partial_{m\beta}S_{lm}' \notag\\
		=& N! [\widetilde{S}_{mn} ie(\partial_{n\alpha}A_n^\beta S_{nm} + A_n^\beta\partial_{n\alpha}S_{nm} - A_m^\alpha\partial_{m\beta}S_{nm}) \notag\\
		&+ \sum_{l\neq n,m} [\widetilde{S}_{mn;nl} ieA_n^\alpha\partial_{m\beta}S_{lm} - \widetilde{S}_{mn;lm} ieA_m^\beta\partial_{n\alpha}S_{ln}] \notag\\
		&+ \sum_{l,l'\neq n,m}\widetilde{S}_{mn;l'l} e^{-i\theta_{l'n}}[\partial_{n\alpha}S_{l'n}-ieA_{l'}^\alpha S_{l'n}] \notag\\
		&\times e^{i\theta_{lm}}[\partial_{m\beta}S_{lm}+ieA_{l}^\beta S_{lm}] ]+\mathcal{O}(B^2) \notag\\
		=& N! \widetilde{S}_{mn} ie(\epsilon^{\alpha\beta\gamma}B^\gamma S_{nm}/2 + A_n^\beta\partial_{n\alpha}S_{nm} - A_m^\alpha\partial_{m\beta}S_{nm}) \notag\\
		&+ \sum_{l\neq n} \widetilde{S}_{mn;nl} ieA_n^\alpha\partial_{m\beta}S_{lm} - \sum_{l'\neq m} \widetilde{S}_{mn;l'm} ieA_m^\beta\partial_{n\alpha}S_{l'n} \notag\\
		&+ \sum_{l,l'\neq n,m}\widetilde{S}_{mn;l'l} \times[e^{i(\theta_{lm}-\theta_{l'n})}\partial_{n\alpha}S_{l'n}\partial_{m\beta}S_{lm}\notag\\
		&-ieA_{l'}^\alpha S_{l'n}\partial_{m\beta}S_{lm}+ieA_{l}^\beta S_{lm}\partial_{n\alpha}S_{l'n}] + \mathcal{O}(B^2)
		\end{align}
		
		\subsection{The reduced form of $ a_\mu $}
		\begin{align}
		&\text{The 1st term of \eqref{eq:BerryConn}} \notag\\
		=& i\frac{-1}{N!(\det S)^2} \sum_{l\neq n} \widetilde{S}_{nl} \partial_{n\alpha} S_{ln} 
		\times N!\det S +\mathcal{O}(B^2) \notag\\
		=& \frac{-i}{\det S}\sum_{l\neq n} \widetilde{S}_{nl} \partial_{n\alpha} S_{ln} 
		+\mathcal{O}(B^2),
		\label{conn_1}
		\end{align}
		\begin{align}
		&\text{The 2nd term of \eqref{eq:BerryConn}} \notag\\
		=& \frac{1}{N!\det S} \times N! \bigl[\sum_l ie A_l^\alpha \widetilde{S}_{nl} S_{ln} 
		+ \sum_{l\neq n} \widetilde{S}_{nl} \partial_{n\alpha} S_{ln}\bigr] +\mathcal{O}(B^2) \notag\\
		=& \frac{1}{\det S} \bigl[\sum_l ie A_l^\alpha \widetilde{S}_{nl} S_{ln} 
		+ \sum_{l\neq n} \widetilde{S}_{nl} \partial_{n\alpha} S_{ln}\bigr] +\mathcal{O}(B^2)
		\label{conn_2}
		\end{align}
		
		\subsection{The reduced form of $ F_{\mu\nu} $}
		\begin{align}
		&(\text{The 1st line of \eqref{eq:BerryCurv}}) \nonumber\\
		=& -2\hbar{\rm Im}~\left[\frac{1}{(\det S)^2} \sum_{l}\widetilde{S}_{nl} \partial_{n\alpha} S_{ln}  \sum_{l'}\widetilde{S}_{ml'} \partial_{m\alpha} S_{l'm}\right] \notag\\
		&+ \mathcal{O}(B^2)\notag\\
		=&0.
		\label{1st}
		\end{align}
		In the 2nd line of \eqref{1st}, the ingredient of the parenthese is real-valued, so that its imaginary part is equal to zero.
		\begin{align}
		&(\text{The 2nd line of \eqref{eq:BerryCurv}})\nonumber\\ 
		=& {\rm Im} \bigl[\frac{2}{(\det S)^2} \sum_{l\neq n} [\widetilde{S}_{nl}\partial_{n\alpha} S_{ln}] \notag\\
		&\times [ieA_m^\beta\widetilde{S}_{mm}+\sum_{l'\neq m} \widetilde{S}_{ml'} (\partial_{m\beta}S_{ml'} +ieA_{l}^\beta  S_{ml'})] \notag\\
		&+ \mathcal{O}(B^2)\bigr] \notag\\
		=& \frac{2}{(\det S)^2}\sum_{l\neq n} [\widetilde{S}_{nl}\partial_{n\alpha} S_{ln}] \notag\\
		& \times [eA_m^\beta\widetilde{S}_{mm}+\sum_{l'\neq m} \widetilde{S}_{ml'}S_{ml'} eA_{l'}^\beta ] + \mathcal{O}(B^2)
		\label{2nd}
		\end{align}
		The 3rd line of \eqref{eq:BerryCurv} can be obtained from \eqref{2nd} by exchanging $ n\alpha $ and $ m\beta $ and taking the complex conjugate:
		\begin{align}
		&(\text{The 3rd line of \eqref{eq:BerryCurv}})\nonumber\\ 
		=&\frac{2}{(\det S)^2} \sum_{l\neq m} [\widetilde{S}_{ml}\partial_{m\beta} S_{lm}] \notag\\
		& \times [eA_n^\alpha\widetilde{S}_{nn}+\sum_{l'\neq n} \widetilde{S}_{nl'}S_{nl'} eA_{l'}^\alpha ] + \mathcal{O}(B^2)
		\label{3rd}
		\end{align}
		For $ n=m $,
		\begin{align}
		&(\text{The 4th line of \eqref{eq:BerryCurv}}) \nonumber\\
		=&{\rm Im}\frac{-2}{|S|} \Bigl[\widetilde{S}_{nn} i\epsilon^{\alpha\beta\gamma}eB^\gamma/2 \notag\\
		&+ \sum_{l,l'\neq n} \widetilde{S}_{nn;l'l} \times [e^{i(\theta_{nl'}-\theta_{ln})} \partial_{n\alpha}S_{nl'}\partial_{n\beta}S_{ln} \notag\\ 
		& -ieA_{l'}^\alpha S_{nl'}\partial_{n\beta}S_{ln} +ieA_{l}^\beta S_{ln}\partial_{n\alpha}S_{nl'}]\Bigr]
		\label{4th_1}
		\end{align}
		For $ n\neq m $,
		\begin{align}
		&(\text{The 4th line of \eqref{eq:BerryCurv}}) \nonumber\\
		=& {\rm Im}\frac{-2}{|S|} \Bigl[\widetilde{S}_{mn} ie[\frac{1}{2}\epsilon^{\alpha\beta\gamma}B^\gamma S_{nm} \notag\\
		&+ A_n^\beta\partial_{n\alpha}S_{nm} - A_m^\alpha\partial_{m\beta}S_{nm}] \notag\\
		&+ \sum_{l\neq n,m} [\widetilde{S}_{mn;nl} ieA_n^\alpha\partial_{m\beta}S_{lm} - \widetilde{S}_{mn;lm} ieA_m^\beta\partial_{n\alpha}S_{ln}] \notag\\
		&+ \sum_{l,l'\neq n,m}\widetilde{S}_{mn;l'l} \times[e^{i(\theta_{lm}-\theta_{l'n})}\partial_{n\alpha}S_{l'n}\partial_{m\beta}S_{lm}\notag\\
		&-ieA_{l'}^\alpha S_{l'n}\partial_{m\beta}S_{lm}+ieA_{l}^\beta S_{lm}\partial_{n\alpha}S_{l'n}]\Bigr]
		\label{4th_2}
		\end{align}
		
		\subsection{Special case: square lattice}
		We now derive the explicit form of $ a_n^\alpha $ and 
		$ F_{mn}^{\alpha\beta} $ of the square lattice 
		under uniform magnetic field in $ z $-direction. 
		Here, we assume that the overlap integral of the nearest-neighbor 
		atoms is dominant. 
		Moreover, due to the translational and rotational symmetry, 
		$ S_{nm} $ takes the same value for all the nearest-neighbor pair $ (n,m) $, 
		and $ \partial_{n\alpha}\mathcal{N} = 0 $. 
		Then, from Eqs.~\eqref{conn_1} and \eqref{conn_2},
		\begin{subequations}
			\begin{align}
			a_{n}^x =& \frac{-eB}{2\det S} \Bigl[-Y_n \widetilde{S}_{nn} -\sum_{e=\pm x,y} Y_{n+e} \widetilde{S}_{n,n+e} S_{n,n+e}\Bigr] \\
			a_{n}^y =& \frac{-eB}{2\det S} \Bigl[X_n \widetilde{S}_{nn} +\sum_{e=\pm x,y} X_{n+e} \widetilde{S}_{n,n+e} S_{n,n+e}\Bigr] \\
			a_n^z =& 0,
			\end{align}
		\end{subequations}
		where $ n=n\pm x(y) $ indicates the neighboring cite of $ n $ in the $ x(y) $-direction. 
		These are also unifiedly written as
		\begin{align}
		a_n^\alpha = \frac{-e}{\det S} 
		\Bigl[A_n^\alpha \widetilde{S}_{nn} + \sum_{l} A_l^\alpha \widetilde{S}_{nl} S_{nl}\Bigr].
		\end{align}
		The curvature at $ R=R^0 $ can be directly calculated from this $ a_n^\alpha $ by
		$ F_{nm}^{\alpha\beta} = \left. (\partial_{n\alpha} a_m^\beta - \partial_{m\beta} a_n^\alpha) \right|_{R \to R^0} $:
		\begin{subequations}
			\begin{align}
			F_{nn}^{\alpha\beta} =& -\frac{\widetilde{S}_{nn}}{\det S} 
			\epsilon^{\alpha\beta\gamma}eB^\gamma 
			\notag\\&-\sum_{l}\frac{\widetilde{S}_{nl}}{\det S}
			\Bigl[-e(A_n^\beta-A_l^\beta)\partial_{n\alpha}S_{nl} \notag\\
			&+ e(A_n^\alpha-A_l^\alpha)\partial_{n\beta}S_{nl}\Bigr] \\
			F_{nm}^{\alpha\beta} =& -\frac{\widetilde{S}_{nm}}{\det S} \Bigl[\epsilon^{\alpha\beta\gamma}eB^\gamma S_{nm} \notag\\
			&+ e(A_n^\beta-A_m^\beta)\partial_{n\alpha}S_{nm} \notag\\
			&- e(A_n^\alpha-A_m^\alpha)\partial_{n\beta}S_{nm}\Bigr]
			\end{align}
		\end{subequations}
		The obtained formula is symmetric for the exchange of $ n $ and $ m $, and also ensures the screening effect, $ \sum_{m} F_{nm}^{\alpha\beta} = -ZeB $ for any $ n $.

\newpage
\section{II. Band structure of phonons}
In this section, we show that $ Z_{\textrm{ph}} =0 $ holds in our system. 
The argument is essentially the same as the one in 
Ref.~\onlinecite{shindou2013topological}, where $ Z_{\textrm{ph}} =0 $ was proven for magnons. 
We generalized it to phonons by using the unified perspective of 
the Berry curvature of bosonic systems.
	
	\subsection{Consistency of the two definitions}
	Here, we will review the general argument given in Ref.~\onlinecite{raghu2008analogs}.
	We consider the system with some degrees of freedom $\{x_i\}$ or $\{b_i,b_i^{\dagger}\}$. Concretely,
	$i$ stands for the spatial index, polarization degrees of freedom, etc. 
	Note that $x_i$ may include both the coordinate $q_{\alpha}$ and the momentum 
	$p_{\alpha}$.
	We assume that the Hamiltonian of the system is given as follows:
	\begin{equation}
	H= \begin{cases}
	\frac{1}{2}\sum_{i,j}\mathcal{B}_{ij}x_ix_j & (\text{real case})\\
	\sum_{i,j}\mathcal{B}_{ij}b^{\dagger}_ib_j & (\text{complex case})
	\end{cases},
	\end{equation}
	where $\mathcal{B}$ is a real symmetric matrix in real case, and is a hermitian matrix in 
		complex case. We also assume that $\mathcal{B}$ is positive semi-definite from the stability condition.
	Furthermore, we need an information of the commutation relation of 
	$\{x_i\}$s and $\{b_i,b_i^{\dagger}\}$. We assume
	\begin{equation}
	[x_i,x_j]=\mathcal{A}_{ij},\quad
	[b_i,b^{\dagger}_j]=\mathcal{A}_{ij}\, ,\quad
	\mathcal{A}_{ij}:= (i\Omega^{-1})_{ij}.
	\end{equation}
	$\mathcal{A}$ is a pure imaginary 
		anti-symmetric matrix in real case,
		and is a hermitian matrix in complex case.
	We further assume that $\mathcal{A}_{ij}$ does
	not depend on $\{x_i\}$s, and not singular (i.e., has a full rank). 
		The case where $\Omega_{ij}$ is singular corresponds to the system with 
		constraints \cite{faddeev1988hamiltonian}, and we will not consider such a case here.
	Then the EoM is given as
	\begin{align}
	i\dot{x}_i=(\mathcal{AB})_{ij}x_j, \quad i\dot{b}_i=(\mathcal{AB})_{ij}b_j \label{sympreal}
	\end{align}
	If the system possesses the translational symmetry, we can Fourier transform Eq.~(\ref{sympreal}) to obtain 
		$i \dot{x}_{\bm{k}}^{\alpha}=(\mathcal{A}_{\bm{k}}\mathcal{B}_{\bm{k}})^{\alpha\beta}x^{\beta}_{\bm{k}}$,
		$i \dot{b}_{\bm{k}}^{\alpha}=(\mathcal{A}_{\bm{k}}\mathcal{B}_{\bm{k}})^{\alpha\beta}b^{\beta}_{\bm{k}}$, where
		$\mathcal{A}_{\bm{k}},\mathcal{B}_{\bm{k}}$ are hermitian matrices which, in the real case, additionally satisfy
		$\mathcal{A}_{\bm{k}}^*=-\mathcal{A}_{-\bm{k}}$ and $\mathcal{B}_{\bm{k}}^*=\mathcal{B}_{-\bm{k}}$ from the (anti-)symmetry.
	
	The ``non-hermitian Hamiltonian'' given in Ref.~\onlinecite{qin2012berry}
	corresponds to $\mathcal{A}_{\bm{k}}\mathcal{B}_{\bm{k}}$. Concretely, in $\{q_{x,\bm{k}},q_{y,\bm{k}},\Pi_{x,\bm{k}},\Pi_{y,\bm{k}}\}$ basis,
	\begin{align}
	&\mathcal{A}_{\bm{k}}=\left(
	\begin{array}{cc}
	0& i I_2\\
	-i I_2& i G_{\bm{k}}\\
	\end{array}
	\right),\quad
	\mathcal{B}_{\bm{k}}=\left(
	\begin{array}{cc}
	D_{\bm{k}}& 0\\
	0& I_2\\
	\end{array}
	\right),\notag\\
	&\mathcal{A}_{\bm{k}}\mathcal{B}_{\bm{k}}=\left(
	\begin{array}{cc}
	0& iI_2\\
	-iD_{\bm{k}}& iG_{\bm{k}}\\
	\end{array}
	\right), \label{shisetting}
	\end{align}
	where $I_2$ represents $2$ by $2$ unit matrix. The corresponding quantity in 
	Ref.~\onlinecite{shindou2013topological} is, in $\{\beta_{\bm{k}},\beta^{\dagger}_{\bm{-k}}\}$ basis,
	\begin{align}
	&\mathcal{A}_{\bm{k}}=\sigma_3,\quad
	\mathcal{B}_{\bm{k}}=H_{\bm{k}}=
	\left(
	\begin{array}{cc}
	a_{\bm{k}}&b_{\bm{k}} \\
	b_{-\bm{k}}^*& a_{-\bm{k}}^* \\
	\end{array}
	\right)
	,\notag\\
	&\mathcal{A}_{\bm{k}}\mathcal{B}_{\bm{k}}=\sigma_3 H_{\bm{k}}. \label{murakamisetting}
	\end{align}
	Diagonalization of $\mathcal{A}_{\bm{k}}\mathcal{B}_{\bm{k}}$ amounts to solving the full problem, since it leads to the diagonal EoM.
	
	Here, we will show that the definition of Berry curvature given in Ref.~\onlinecite{qin2012berry}
	is equivalent in the general setting given above, and the only difference resides in the form of
	$\mathcal{A}_{\bm{k}}$. Here we will write $\mathcal{A}_{\bm{k}}\mathcal{B}_{\bm{k}}$ 
	as $\tilde{H}_{\bm{k}}$, following Ref.~\onlinecite{qin2012berry}.
	Then, if $\mathcal{B}_{\bm{k}}$ is positive definite, we can consider the following similarity 
	transformation:
	\begin{equation}
	\tilde{H}_{\bm{k}} \to \mathcal{B}_{\bm{k}}^{1/2}\tilde{H}_{\bm{k}}\mathcal{B}_{\bm{k}}^{-1/2}
	= \mathcal{B}_{\bm{k}}^{1/2}\mathcal{A}_{\bm{k}}\mathcal{B}_{\bm{k}}^{1/2},
	\end{equation}
	which is hermitian, so $\tilde{H}_{\bm{k}}$ can be diagonalized. Therefore for positive definite
	$\mathcal{B}_{\bm{k}}$, $\tilde{H}_{\bm{k}}$ is quasi-hermitian. From
	$\tilde{H}_{\bm{k}}^{\dagger}=\mathcal{B}_{\bm{k}}\tilde{H}_{\bm{k}}\mathcal{B}^{-1}_{\bm{k}}$, 
	it is also pseudo-hermitian.
	However, if there is a zero eigenvalue
	of $\mathcal{B}_{\bm{k}}$, then we cannot construct $\mathcal{B}_{\bm{k}}^{-1/2}$, 
	so the above argument fails. 
	This failure of the diagonalization is known in the literature as the defective point or 
	non-hermitian degeneracy \cite{berry2004physics}. 
	The characteristic feature of a defective point is the square root dispersion near the degeneracy,
	and at $\bm{k}=0$ in the phonon system, it leads to the linear spectrum $\omega_{\bm{k}}=c\sqrt{k^2}$
	even though the dynamical matrix is quadratic in $k^\alpha$.

	So, except for the defective point, $\tilde{H}_{\bm{k}}$ can be diagonalized by the matrix
	$T_{\bm{k}}$, which is not necessarily unitary:
	\begin{align}
	&T_{\bm{k}}^{-1}\tilde{H}_{\bm{k}}T_{\bm{k}}=
	\begin{bmatrix}
	\omega_1 &  & \\
	& \ddots & \\
	& &\omega_N 
	\end{bmatrix}
	, \notag\\
	&T_{\bm{k}}=\left(\ket{u_1}\dots\ket{u_N}\right), \label{eomdiag}
	\end{align}
	where $\ket{u_i}$ is the $ N $ dimensional column vector, and is the $i$th eigenvector. Then
	the Berry connection is defined as
	\begin{equation}
	a_{\mu}^{i}={\rm Re} \, (i \braket{u_i|\mathcal{B}_{\bm{k}}\partial_{k_{\mu}}|u_i}), \label{shiberry}
	\end{equation}
	where we imposed the normalization:
	\begin{equation}
	\braket{u_i|\mathcal{B}_{\bm{k}}|u_j}=\delta_{ij}, \quad \bra{u_i}=(\ket{u_i})^{\dagger}. \label{biortho}
	\end{equation}
	We note that, given the right eigenvector of $\tilde{H}_{\bm{k}}=\mathcal{A}_{\bm{k}}\mathcal{B}_{\bm{k}}$ as
		$\ket{u_i}$, we can construct the left eigenvector as $\mathcal{B}_{\bm{k}}\ket{u_i}$, since 
		\begin{equation}
		\tilde{H}_{\bm{k}}^{\dagger}(\mathcal{B}_{\bm{k}}\ket{u_i})=\mathcal{B}_{\bm{k}}\mathcal{A}_{\bm{k}}\mathcal{B}_{\bm{k}}\ket{u_i}
		=\epsilon_i \mathcal{B}_{\bm{k}}\ket{u_i},
		\end{equation}
		where we used the hermicity of $\mathcal{A}_{\bm{k}}$ and $\mathcal{B}_{\bm{k}}$.
		So the normalization (\ref{biortho}) amounts to choosing the biorthogonal basis.
	
	On the other hand, the definition of Berry connection in Ref.~\onlinecite{shindou2013topological}
	is given as:
	\begin{equation}
	a^i_{\mu}=i {\rm Tr}\,[\bm{\Gamma}_i \sigma_3 T_{\bm{k}}^{\dagger} \sigma_3(\partial_{k_\mu}T_{\bm{k}}) ], \label{bermurakami}
	\end{equation}
	where $\bm{\Gamma}_i$ is a matrix taking $+1$ for the $j$th diagonal component and zero otherwise.
	We note that $T_{\bm{k}}$, which is the paraunitary matrix which diagonalizes $H_{\bm{k}}$,
	($T_{\bm{k}}^{\dagger} H_{\bm{k}}T_{\bm{k}}=d_{\bm{k}}$, where $d_{\bm{k}}$ is a diagonal matrix) is
	the same matrix as $T_{\bm{k}}$ in Eq.~(\ref{eomdiag}), since
	\begin{equation}
	T_{\bm{k}}^{-1}\tilde{H}_{\bm{k}}T_{\bm{k}}=T_{\bm{k}}^{-1}\sigma_3 H_{\bm{k}}T_{\bm{k}}
	=\sigma_3 T_{\bm{k}}^{\dagger} H_{\bm{k}}
	T_{\bm{k}}=\sigma_3 d_{\bm{k}},
	\end{equation}
	so it also diagonalizes $\tilde{H}_{\bm{k}}$, where we used the paraunitarity of $T_{\bm{k}}$.
	Noting that, from the biorthogonality (\ref{biortho}),
	\begin{equation}
	T_{\bm{k}}^{-1}=\left(
	\begin{array}{c}
	\bra{u_1}H_{\bm{k}}\\
	\vdots\\
	\bra{u_N}H_{\bm{k}}\\
	\end{array}
	\right),
	\end{equation}
	then, from Eq.~(\ref{bermurakami}),
	\begin{align}
	a^i_{\mu}=i {\rm Tr}\,[\bm{\Gamma}_i T_{\bm{k}}^{-1}(\partial_{k_\mu}T_{\bm{k}}) ]
	=&i (T_{\bm{k}}^{-1}(\partial_{k_\mu}T_{\bm{k}}) )_{ii}
	\notag\\
	=&i\braket{u_i|H_{\bm{k}}\partial_{k_{\mu}}|u_i}.
	\end{align}
	So if we properly identify the matrices $\mathcal{A},\mathcal{B}$, these two expressions are consistent.
	Moreover, we can show that the Berry connection (\ref{bermurakami}) is real by using the fact
	that $[\bm{\Gamma}_i,\sigma_3]=0$:
	\begin{align*}
	(a^{i}_{\mu})^*=&-i {\rm Tr}\,[ (\partial_{k_\mu}T^{\dagger}_{\bm{k}}) \sigma_3 T_{\bm{k}}\sigma_3 \bm{\Gamma}_i]\\
	=&i {\rm Tr}\,[ T^{\dagger}_{\bm{k}} \sigma_3 \partial_{k_\mu} T_{\bm{k}}\sigma_3 \bm{\Gamma}_i]\\
	=&i {\rm Tr}\,[\sigma_3 \bm{\Gamma}_i T^{\dagger}_{\bm{k}} \sigma_3 \partial_{k_\mu} T_{\bm{k}} ]\\
	=&i {\rm Tr}\,[\bm{\Gamma}_i \sigma_3  T^{\dagger}_{\bm{k}} \sigma_3 \partial_{k_\mu} T_{\bm{k}} ]\\
	=&a^{i}_{\mu}.
	\end{align*}
	In general, the Berry connection is complex and we need to take the
	real part of it as is done in Eq.~(\ref{shiberry}) and as is pointed out in Ref.~\onlinecite{raghu2008analogs}.
	
	Although the Berry connection in phonon system is completely
	given as Eq.~(\ref{shiberry}), we can consider the following step-by-step transformation to compare 
	to the definition in the magnon system. 
	First, we diagonalize $\mathcal{A}_{\bm{k}}$ by the unitary matrix $U_{\bm{k}}$:
	\begin{equation}
	\tilde{H}_{\bm{k}} \to U_{\bm{k}}^{\dagger}\mathcal{A}_{\bm{k}}\mathcal{B}_{\bm{k}}U_{\bm{k}}=
	\mathcal{D}_{\bm{k}}U^{\dagger}_{\bm{k}}\mathcal{B}_{\bm{k}}U_{\bm{k}}:= \mathcal{D}_{\bm{k}}\tilde{\mathcal{B}}_{\bm{k}},
	\end{equation}
	where $\mathcal{D}_{\bm{k}}$ is the diagonal matrix with both positive and negative element, but does not
	have a zero component from the assumption of nonsingularity. 
	Then we consider the following transformation:
	\begin{align}
	\mathcal{D}_{\bm{k}}\tilde{\mathcal{B}}_{\bm{k}} \to& {(\mathcal{D}_{\bm{k}}^{+}})^{-\frac{1}{2}}\mathcal{D}_{\bm{k}}\tilde{\mathcal{B}}_{\bm{k}}{(\mathcal{D}_{\bm{k}}^{+}})^{\frac{1}{2}}\notag\\
	=&({(\mathcal{D}_{\bm{k}}^{+}})^{-1}\mathcal{D}_{\bm{k}})({(\mathcal{D}_{\bm{k}}^{+}})^{\frac{1}{2}}\tilde{\mathcal{B}}_{\bm{k}}{(\mathcal{D}_{\bm{k}}^{+}})^{\frac{1}{2}}) \notag\\
	=&: \omega \mathcal{B}_{\bm{k}}',
	\end{align}
	where $\omega$ is the diagonal matrix with $p$ numbers of
	$+1$ components and $q$ numbers of $-1$ components, and the superscript ``$ + $" here indicates taking the absolute value of each component. In the case of 
	Eq.~(\ref{shisetting}), $p=q$, so $\omega=\sigma_3$. Therefore, the problem has been reduced to
	the same structure as Eq.~(\ref{murakamisetting}). In general, we consider the matrix 
	$T_{\bm{k}} \in SU(p,q)$ which diagonalizes 
	$\mathcal{B}_{\bm{k}}' \to T_{\bm{k}}^{\dagger}\mathcal{B}_{\bm{k}}'T_{\bm{k}}=\mathcal{D}_{\bm{k}}'$. Then the Berry connection
	is given as
	\begin{align}
	&a^i_{\mu}\notag\\
	=&\text{Re}(i {\rm Tr}\,[\bm{\Gamma}_i (U_{\bm{k}}{(\mathcal{D}_{\bm{k}}^{+}})^{\frac{1}{2}}T_{\bm{k}})^{-1}\partial_{k_\mu}
	(U_{\bm{k}}{(\mathcal{D}_{\bm{k}}^{+}})^{\frac{1}{2}}T_{\bm{k}}) ])\\
	=&
	\text{Re}(i {\rm Tr}\,[\bm{\Gamma}_i \omega T^{\dagger}_{\bm{k}}\omega
	{(\mathcal{D}_{\bm{k}}^{+}})^{-\frac{1}{2}}U_{\bm{k}}^{\dagger}
	\partial_{k_\mu}
	(U_{\bm{k}}{(\mathcal{D}_{\bm{k}}^{+}})^{\frac{1}{2}}T_{\bm{k}}) ]).
	\end{align}
	
	\section{Sum rule of Chern number for positive frequency bands}
	In Ref.~\onlinecite{shindou2013topological}, the authors showed that the sum of Chern numbers
 	of positive frequency bands is zero. Here we will show that the same result holds for 
	Eq.~(\ref{shisetting}) with positive definite $D_{\bm{k}}$.
	First, we construct the adiabatic path $\mathcal{B}_{\bm{k},\lambda}$ and $\tilde{H}_{\bm{k},\lambda}$ as 
	$\mathcal{B}_{\bm{k},\lambda}=(1-\lambda)\mathcal{B}_{\bm{k}}+\lambda I_4$ and 
	$\tilde{H}_{\bm{k},\lambda}=\mathcal{A}_{\bm{k}}\mathcal{B}_{\bm{k},\lambda}$ $(\lambda \in [0,1])$.
	We observe that,
		\begin{equation}
		\text{det}\,(\tilde{H}_{\bm{k},\lambda})=\text{det}\,(\mathcal{A}_{\bm{k}})\text{det}\,(\mathcal{B}_{\bm{k},\lambda})
		=\text{det}\,(D_{\bm{k},\lambda}),
		\end{equation}
		where $D_{\bm{k},\lambda}=(1-\lambda)D_{\bm{k}}+\lambda I_2$.
		So the spectrum of $\tilde{H}_{\bm{k},\lambda}$ never crosses 
		zero, since $D_{\bm{k},\lambda}$ is positive definite.
		Therefore, we can deform $\tilde{H}_{\bm{k}}$ to $\mathcal{A}_{\bm{k}}$ without closing the gap
		between positive and negative energy subspaces.
	
	Then the sum of Chern numbers of positive frequency bands for
	$\tilde{H}_{\bm{k}}$ is the same as that of $\mathcal{A}_{\bm{k}}$. So the remaining problem is
	to determine the sum of Chern numbers of $\mathcal{A}_{\bm{k}}$. Here,
	\begin{align}
	\mathcal{A}_{\bm{k}}=&\left(
	\begin{array}{cc}
	0& i I_2\\
	-i I_2& g(k)\sigma_2\\
	\end{array}
	\right)\notag\\
	=&I_2\otimes s_2+g(\bm{k})\sigma_2\otimes\left(\frac{I_2-s_3}{2}\right) 
	\end{align}
	By rotating $\sigma_2 \to \sigma_3$ and interchanging the basis $s_i \leftrightarrow \sigma_i$,
	\begin{align}
	\mathcal{A}_{\bm{k}}=&\sigma_2\otimes I_2+g(\bm{k})\left(\frac{I_2-\sigma_3}{2}\right) \otimes s_3
	\notag\\
	=&\left(
	\begin{array}{cc}
	\sigma_2+g(\bm{k})\left(\frac{I_2-\sigma_3}{2}\right)& 0\\
	0& \sigma_2-g(\bm{k})\left(\frac{I_2-\sigma_3}{2}\right)\\
	\end{array}
	\right),
	\end{align}
	so the system is block diagonalized into two sectors. However, each sector only contains two Pauli
	matrices $\sigma_2$ and $\sigma_3$, so the Chern number is identically zero for each band.
	Therefore, we showed that the sum of Chern numbers of positive frequency 
	bands are zero for (\ref{shisetting}).